\documentclass[reprint,superscriptaddress]{revtex4-2}
\usepackage[utf8]{inputenc}
\usepackage{amsmath,bm}
\usepackage{amssymb}
\usepackage{graphicx}
\usepackage{import}
\usepackage{color}
\usepackage{tikz}
\usepackage{makecell}
\usepackage{booktabs}
\usepackage{placeins}
\usepackage{chngcntr}
\usepackage{setspace}
\usepackage{parskip}
\usepackage{hyperref}
\usepackage{cleveref}
\crefname{equation}{equation}{equations}
\crefname{figure}{Fig.}{Figs.}
\crefname{tabular}{Table}{Tables}
\crefname{supfig}{Extended Data Fig.}{Extended Data Figs.}

\graphicspath{{Images/}{figs/}}

\newcommand{\Dp}{D_\mathrm{p}}

\newcommand{\mup}{\mu^\mathrm{(p)}}
\newcommand{\mus}{\mu^\mathrm{(s)}}

\begin{document}

\title{Self-organization of primitive metabolic cycles due to non-reciprocal interactions} %{Self-organizing catalytic cycles}

\author{Vincent Ouazan-Reboul}
\affiliation{Max Planck Institute for Dynamics and Self-Organization, Am Fassberg 17, D-37077, G\"{o}ttingen, Germany}
\author{Jaime Agudo-Canalejo}
\affiliation{Max Planck Institute for Dynamics and Self-Organization, Am Fassberg 17, D-37077, G\"{o}ttingen, Germany}
\author{Ramin Golestanian}
\affiliation{Max Planck Institute for Dynamics and Self-Organization, Am Fassberg 17, D-37077, G\"{o}ttingen, Germany}
\affiliation{Rudolf Peierls Centre for Theoretical Physics, University of Oxford, OX1 3PU, Oxford, UK}

\begin{abstract}
We study analytically and numerically a model metabolic cycle composed of an arbitrary number of species of catalytically active particles.
Each species converts a substrate into a product, the latter being used as the substrate by the next species in the cycle.
Through a combination of catalytic activity and chemotactic mobility, the active particles develop effective non-reciprocal interactions with particles belonging to neighbouring species in the cycle.
We find that such model metabolic cycles are able to self-organize through a macroscopic instability, with a strong dependence on the number of species they incorporate.
The parity of that number has a key influence: cycles containing an even number of species are able to minimize repulsion between their component particles by aggregating all even-numbered species in one cluster, and all odd-numbered species in another.
Such a grouping is not possible if the cycle contains an odd number of species, which can lead to oscillatory steady states in the case of chasing interactions.

% One of the greatest mysteries concerning the origin of life is how it has emerged so quickly after the formation of the earth. In particular, it is not understood how the intricate structures of metabolic cycles, which power the non-equilibrium activity of cells and support their functions under homeostatic conditions, have come into existence in the first instances. These structures have emerged from a dilute primordial soup of chemicals that have turned out to be suitable partners in certain reactions in the roles of reactants and catalysts. While it is generally expected that non-equilibrium conditions would have been necessary for the formation of these primitive metabolic structures, the focus has so far been on externally imposed non-equilibrium conditions, such as temperature or proton gradients. Here, we propose an alternative paradigm in which naturally occurring non-reciprocal interactions between catalysts that can potentially partner together in a cyclic reaction lead to their rapid recruitment into self-organized functional structures. We uncover different classes of self-organized cycles that form through exponentially rapid coarsening processes, depending on the parity of the cycle and the nature of the interaction motifs, which are all generic but have readily tuneable features. Our results also shed light on possibilities that may be explored in designing efficient synthetic cycles. %208 words
\end{abstract}

\maketitle

%\section*{Abstract}

%\section*{Main Text}
\label{sec:main}

\begin{figure*}[t]
    % \centering
    \includegraphics[width=0.85\textwidth]{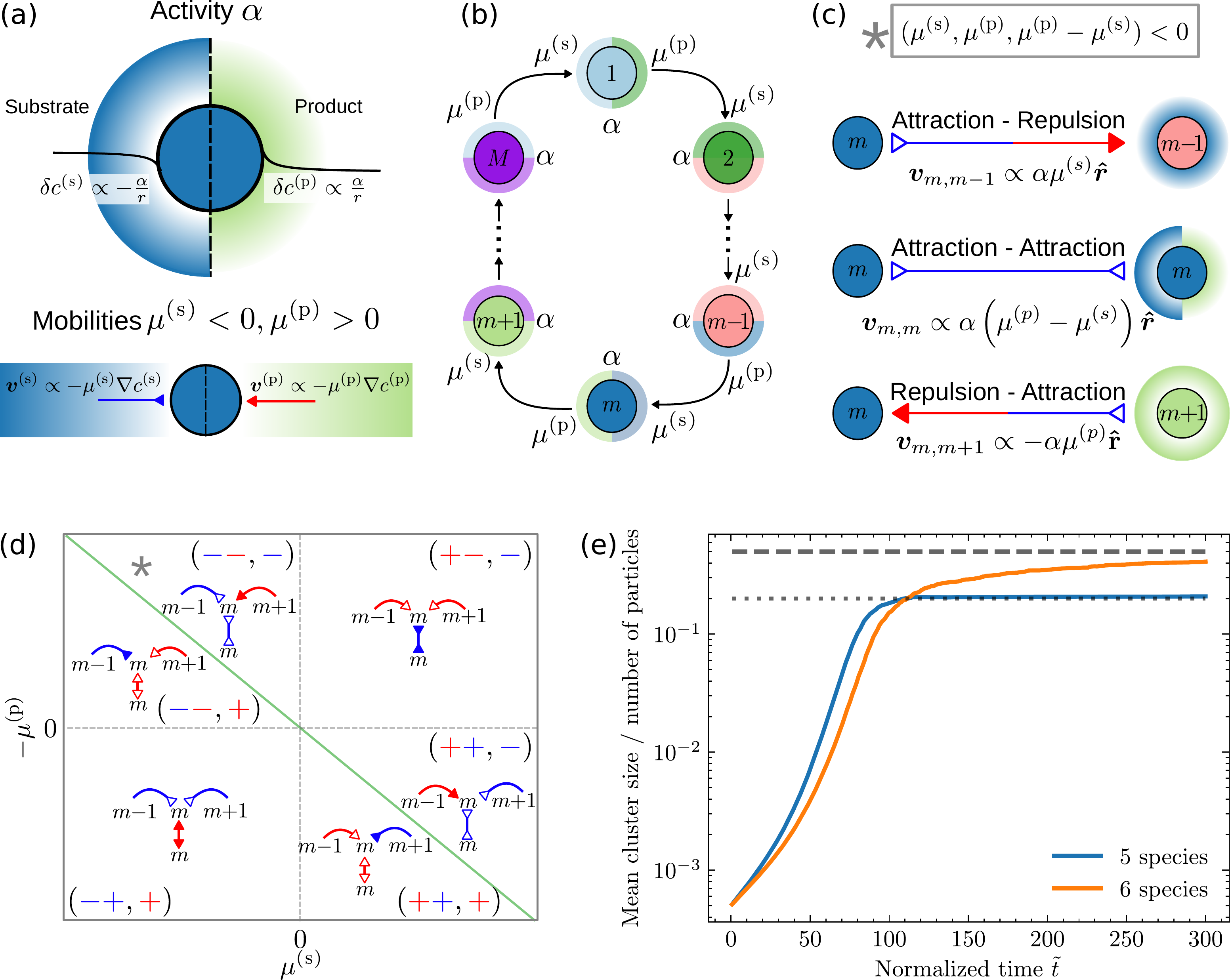}
    \caption{        {Properties and interactions of catalytically-active particles.}
       Panel (a) describes the properties of the studied active particles, which convert substrate (s) into product (p) with a rate given by the activity $\alpha$ (top) and respond to gradients of these chemicals with mobilities $\mus$ and $\mup$ (bottom).
        Panel (b) illustrates our system, which $M$ active particle species are arranged in a model catalytic cycle, in which the product of species $m$ is the substrate of species $m+1$.
        Panel (c) shows how non-reciprocal interactions arise between particles of the same or adjacent species. The direction and colour of arrows indicate the attractive (blue, inwards arrowhead) or repulsive (red, outwards arrowhead) nature of the interaction.
        Panel (d) contains the phase diagram of interaction motifs. Each region constrains the mobilities so that one interaction has a higher magnitude than the others, as highlighted by a full arrowhead. The grey asterisk indicates the location in parameter space of the interactions pictured in panel (c). The green line separates the self-attracting and self-repelling regions. The sign triplets correspond to the signs of $(\mus,\mup,\mup-\mus)$.
        Panel (e) gives the cluster growth dynamics for a cycle of $M=5$ (blue, see \cref{fig:even_odd_aggreg}(a)) and Movie 7)
        and $M=6$ (orange, see \cref{fig:even_odd_aggreg}(b) and Movie 3)
        species, showing super-exponential coarsening dynamics of the condensate formation.
        For $ M = 5 $ species, the mean cluster size saturates at a value corresponding to the total particle population divided by $ M $ (grey dotted line).
        For $ M = 6 $, it saturates at half the particle population (dashed grey line).
}
    \label{fig:parts_cycles}
\end{figure*}

Since Oparin \cite{oparin1938Origin} proposed a picture to describe how early forms of living matter might have emerged from what Haldane described as the prebiotic soup \cite{haldane1957}, there has been a significant amount of progress in our understanding of the physical aspects of the origin of life. Recent examples of such studies include spontaneous emergence of catalytic cycles  \cite{ZoranaRMP,zeravcic2017Spontaneous}, spontaneous growth and division of chemically active droplets \cite{zwicker2017Growth,tena2021accelerated,matsuo2021Proliferating}, programmable self-organization of functional structures under non-equilibrium conditions \cite{McMullen2022,Osat2022}, and controllable realization of metabolically active condensates \cite{testa2021Sustained}. A striking generic observation that has emerged in a variety of different scenarios is that the introduction of non-equilibrium activity in the form of catalytic activity, or a primitive form of metabolism, can be a versatile driving force for functional structure formation \cite{Soto2014,agudo-canalejo2018Phoresis,giunta2020Crossdiffusion,Cotton:2022} with manifestations of lifelike behaviour \cite{RG-prl-2012,Palacci2013,Cohen2014,Stark2014,Saha2014,Soto2015,agudo-canalejo2019Active,qiao2017Predatory,meredith2020Predator}. It has also been demonstrated that the structured catalytic activity that would support the required non-equilibrium processes for primitive cells can be successfully coupled with the condensation of appropriate functional nucleotide and peptide components in membrane-free systems \cite{koga2011Peptide,sokolova2013Enhanced,drobot2018Compartmentalised}, as well as lipid components in protocells with functionalized membranes \cite{Kamat2015,Jin2018}. 

The physicochemically motivated ideas initiated by Oparin and Haldane were critically debated for much of the past century by proponents of the perspective that (genetic) information should be considered as the main organizer of matter that forms life \cite{Lazcano:2012}. As a modern interpretation of these considerations, one can pose the following question: how can we envisage pathways in which the information contained in chemical reaction networks from which primitive forms of metabolism can emerge would lead to structural self-organization of the corresponding components? Here, we propose a strategy that can achieve this task by employing the naturally occurring non-reciprocal interactions between catalysts that can form a cyclic reaction network. We show that model catalytically-active particles participating in a metabolic cycle 
are able to spontaneously self-organize into condensates, which may aggregate or separate
depending on the number of particle species involved in the cycle, and exhibit chasing, periodic aggregation and dispersal, as well as self-stirring, thus providing a generic mechanism for spontaneous formation of metabolically-active protocells.

Non-reciprocal interactions have been shown to generically emerge in active matter in the context of non-equilibrium phoretic interactions between catalytically active colloids and enzymes ~\cite{Ramin-LesHouches2018}. Let us consider a set of $M$ species of chemically-active particles (\cref{fig:parts_cycles}(a), top), representing catalyst molecules or enzymes. Each of the particles converts a substrate (s) into a product (p) at a rate $\alpha$. At steady state, they create perturbations in the concentration field of the corresponding substrate that decays with distance $r$ as $\delta c^\mathrm{(s)} \propto - \alpha / r$, and a corresponding change in the concentration of the corresponding product as $\delta c^\mathrm{(p)} \propto \alpha / r$. %(Methods).
These particles are also chemotactic (\cref{fig:parts_cycles}(a), bottom):
when subjected to a concentration gradient of their substrate, they develop a velocity 
$\bm{v} \propto - \mus \nabla c^{\mathrm{(s)}}$ with  $\mus$ the chemotactic mobility for the substrate, which is negative or positive if the particle is attracted to or repelled from the substrate, respectively. Similarly, the particles are able to chemotax in response to gradients of their products, with a mobility $\mup$.

To create a model for primitive metabolism, we consider a simplified catalytic cycle (\cref{fig:parts_cycles}(b)), in which the substrate of the catalyst species $m$,
which we denote as chemical $(m)$, is the product of species $m-1$. To close the cycle, species $1$ has the product of species $M$ as its substrate. For simplicity, we take all catalyst species to have the same parameters $\alpha$, $\mus$ and $\mup$, and to be present in the system at identical initial concentrations. The cycle can achieve a steady state without net chemical production or consumption. Due to their chemical activity and chemotactic mobilities, the particle species can interact with one another through chemical fields (\cref{fig:parts_cycles}(c)). For instance, if we consider two particles of species $m$ and $m-1$, then the particle of species $m-1$ creates, through its chemical activity, a concentration gradient of the substrate of the particle of species $m$, to which the latter responds by developing a velocity directed towards the particle of species $m-1$, $\bm{v}_{m, m-1} \propto \alpha \mus \hat{\bm{r}}$, where $\hat{\bm{r}}$ is the unit vector pointing from the particle that creates the perturbation to the particle that responds to the perturbation
(see Supplementary Information). 
On the other hand, the particle of species $m$ consumes the product of $m-1$, and thus the particle of species $m-1$ develops a velocity $\bm{v}_{m-1,m} \propto - \alpha \mup \hat{\bm{r}}$ towards the other particle. As a consequence, the interactions between the particles of species $m$ and $m-1$ are non-reciprocal, i.e.~$\bm{v}_{m,m-1} \neq - \bm{v}_{m-1,m}$ (see \cref{fig:parts_cycles}(d) for different possibilities). This effective violation of action-reaction symmetry is a signature of non-equilibrium activity, leading to non-trivial many-body behaviour as has been shown for chemically-active particles interacting through a single chemical \cite{agudo-canalejo2019Active,ouazan-reboul2021Nonequilibrium}, active mixtures interacting through generic short-range interactions \cite{saha2020Scalar,you2020Nonreciprocity}, complex plasmas \cite{ivlev2015Statistical}, and other systems \cite{Loos2020,fruchart2021Nonreciprocal}. Particles of the same species also self-interact by consumption of their substrate and creation of their product, with a velocity $\bm{v}_{m,m} \propto \alpha(\mup - \mus) \hat{\bm{r}}$. We note that these effective non-reciprocal interactions mediated by chemical fields are long-ranged, with the induced velocities going as $1/r^2$. %(see Supplementary Information).

We consider the evolution equations for the concentration fields of the active species
$\rho_m$ and their substrates $c^{(m)}$, given by the coupled system of $2M$ equations
\begin{subequations}
	\label{eq:evol}
	\begin{equation}
		\label{eq:rho_i_time_ev}
        \partial_t \rho_m (\bm{r}, t)=\nabla \cdot [\Dp \nabla \rho_m + (\mus \nabla c^{(m)} + \mup \nabla c^{(m+1)}) \rho_m],
    \end{equation}
	\begin{equation}
		\label{eq:c_time_ev}
        \partial_t c^{(m)} (\bm{r}, t) = D \nabla^2 c^{(m)} + \alpha \left(\rho_{m-1} - \rho_m \right).
	\end{equation}
\end{subequations}
Equation \eqref{eq:rho_i_time_ev} describes the conserved dynamics of the catalysts, with a diffusion term involving a species-independent coefficient $\Dp$ and a chemotactic drift term in response to both substrate and product gradients. The substrate concentrations evolve according to the reaction-diffusion \cref{eq:c_time_ev}, with a diffusion coefficient $D$, and a reaction term corresponding to the activity of the catalysts.

\begin{figure}[t]
    % \centering
    \includegraphics[width=\columnwidth]{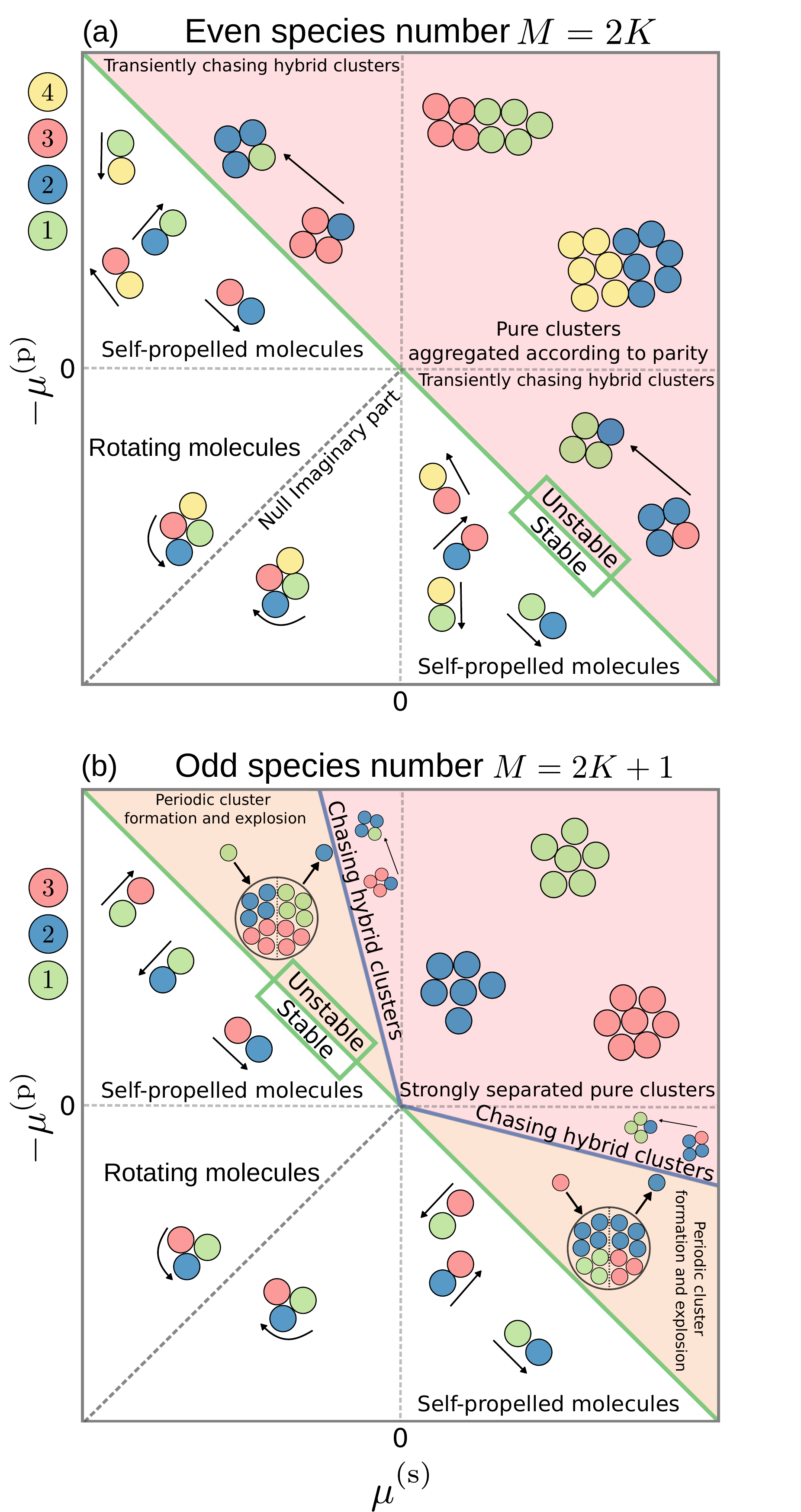}
    \caption{
        {Stability diagrams for catalytic cycles.}
         The cycles contain an even, in panel (a), or an odd, in panel (b), number of catalytic species. The interaction motifs in each quadrant of the parameter space are the same as those displayed in panel (d) of \cref{fig:parts_cycles}. Details of the behaviour in each phase are given in the text. In both the self-propelled and rotating molecule phases, the molecules exchange particles with one another. Molecule rotation does not occur along the dashed lines corresponding to null imaginary part in the unstable eigenvalue.
    }
    \label{fig:phase_diag}
\end{figure}

The time evolution of \cref{eq:evol} naturally leads to the formation of clusters, akin to active phase separation \cite{agudo-canalejo2019Active,ouazan-reboul2021Nonequilibrium}. The clusters are formed through a particularly fast and efficient coarsening process that exhibits exponential growth rather than the commonly occurring power law form, associated with processes such as Ostwald ripening, as can be seen in \cref{fig:parts_cycles}(e)(see Supplementary Information).
This behaviour can be characterized using a simple scaling argument. When particles are collapsing onto a cluster, the rate of growth for the cluster can be estimated as $\frac{d N}{d t}=\oint_S \rho {\bm v} \cdot d {\bm S} $ where the velocity ${\bm v}=-\mu \nabla c$ can be expressed in terms of the particle concentration by using Gauss theorem and the relation $-\nabla^2 c=\alpha \rho/D$, which yields $\frac{d N}{d t}=\frac{\mu \alpha}{D}\rho N$. This expression can be integrated to obtain $N(t)=N_0 \exp\left(\frac{\mu \alpha}{D} \int_{0}^t d t_1 \rho\right)\simeq N_0 \exp\left(\frac{\mu \alpha}{D} \rho t\right)$, which predicts an exponential growth law for constant $\rho$ and allows for super-exponential growth if the density increases with time, which matches well with the results presented in \cref{fig:parts_cycles}(e). This observation suggests that non-equilibrium phoretic interactions have the ability to guide formation of dense clusters in a fast and efficient manner, and as such, can be strong candidates for creating the appropriate conditions for the emergence of early functionalized protocells.

\begin{figure*}[t]
    \centering
    \includegraphics[width=\textwidth]{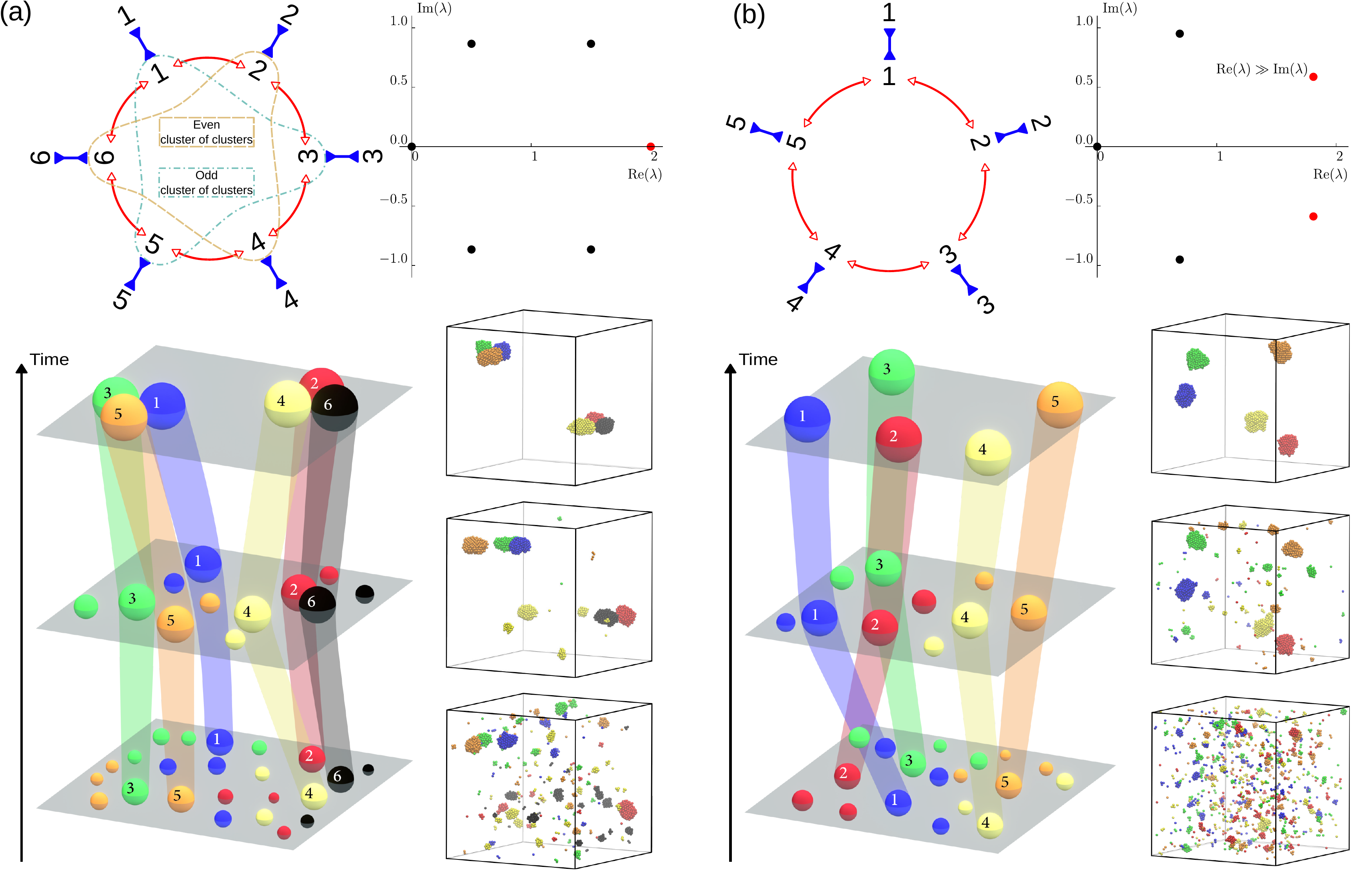}
    \caption{
        {Aggregation dynamics for self-attracting, cross-repelling species.}
        The cycles contain respectively an even or an odd  number of catalytic species in panels (a) and (b). Top left: interactions between the active species. All species are self-attracting and repel both neighbours in the cycle (corresponding to the top-right quadrants in \cref{fig:phase_diag} panels (a) and (b)). Top right: corresponding eigenvalue spectrum, in units of $-(\mup{} - \mus{}) \alpha \rho_0/D$ for the real part and $-(\mus{} + \mup{}) \alpha \rho_0/D$ for the imaginary part. The eigenvalue (or complex conjugate pair) with the largest real part is shown in red. In panel (b), the corresponding conjugate pair has an imaginary part much smaller than its real part, so that the dynamics of the system are non-oscillatory. Bottom: Schematic representations of the time evolution of the aggregation dynamics (left) and corresponding snapshots of molecular dynamics simulations (right)%, see Movies 3 and 7 for even and odd cases, respectively)
        are shown side by side. Dashed lines in panel (a) indicate the parity-based aggregation that occurs for an even number of species.
}
    \label{fig:even_odd_aggreg}
\end{figure*}

A linear stability analysis on \cref{eq:evol} %(Supplementary Information) 
around a spatially-homogeneous solution
%$\{ \rhozrm}, c^{(m)}_H \}$, 
%with perturbations as Fourier modes of wavenumber $q$, leading
leads to the following eigenvalue equation for the macroscopic (long-wavelength) particle density modes
%\begin{equation}
%    \label{eq:stability_eq}
%    - \sum\limits_{n=1}^{M}\eta_{mn} \delta \rho_n = [\lambda + D q^2]\delta \rho_m
%\end{equation}
\begin{equation}
	\label{eq:stability_eq}
	- \sum\limits_{n=1}^{M}\Lambda_{m,n} \delta \rho_n = \lambda\delta \rho_m.
\end{equation}
The matrix $\Lambda_{m,n}$ describes the velocity response of species $m$ to species $n$, and is defined as follows
\begin{equation}
    \label{eq:full_ints}
    \begin{cases}
        \Lambda_{m, m-1} =  \alpha \mus \rho_0/D, \\
        \Lambda_{m,m} = \alpha \left(\mup-\mus\right) \rho_0/D,\\
        \Lambda_{m, m+1} = - \alpha \mup \rho_0/D,\\
        \Lambda_{m, n\notin \left\{ m, m\pm1 \right\}} = 0,
    \end{cases}
\end{equation}
where $\rho_0$ represents the initial homogeneous concentrations. By definition, $\Lambda_{m,n}$ is negative, or positive, if $m$ is attracted to, or repelled from, $n$, respectively.
%The self-interaction for a given species in \cref{eq:full_ints} 
%is a linear combination of its
%cross-interactions with its neighbours in the cycle, leading to
The form of $\Lambda_{m,n}$ suggests a classification scheme as there are six possible interaction motifs (\cref{fig:parts_cycles}(d)), representing the interactions of each species with itself as well as with its two neighbours in the catalytic cycle. The signs of the interactions are represented diagrammatically, following the conventions defined in \cref{fig:parts_cycles}(c) and (d).

The eigenvalues $\lambda_\ell$ ($\ell \in \{ 1, \ldots, M \}$) allow us to predict the stability of the system: 
%Here, the sign of the eigenvalue $\lambda$ predicts the stability of the system: 
$\operatorname{Re} ( \lambda ) > 0$ for any eigenvalue $\lambda$ indicates an instability, whereas $\operatorname{Re} (\lambda)< 0$ for all eigenvalues implies a stable homogeneous state. The eigenvector $\delta \rho^{\ell}_m$, in turn, gives the stoichiometry at the onset of instability, i.e.~the ratio of the different species within the growing perturbation, which may be positive, for species that aggregate together, or negative, for species that separate.
%Since we only seek to determine whether the system is unstable or not,
%we will focus on the $q^2 =0$ mode, which is the most unstable and corresponds to the largest wave length, i.e.~to system-wide, macroscopic instability. It can be shown that the system always has one zero eigenvalue, denoted as $\lambda_0 = 0$,
%and $M-1$ non-null eigenvalues, of which only one needs to be positive in order 
%for an instability to occur.

%What differences are there between $M$ species of active particles 
%interacting only through a single chemical they either produce or consume,
%and $M$ species in a catalytic cycle?
The topology of the catalytic cycle strongly influences its self-organization. As a point of comparison, we consider a non-cyclic system, %\cite{agudo-canalejo2019Active}, 
in which $M$ catalytic species all act on a single chemical field. In this case, all the coefficients of the interaction matrix are equal to $\alpha\mu\rho_{0}$, leading to a system with only one nonzero eigenvalue $\lambda = - M \alpha\mu\rho_{0}/D$. The corresponding instability condition is $\alpha\mu < 0$, % the particles have to be self-attracting, 
and the instability is equivalent to the Keller-Segel model 
% \cite{agudo19,kell70}.
\cite{agudo-canalejo2019Active,keller1970Initiation}.
The model catalytic cycle studied here, however, presents a different category.
%In order to determine which cycles are unstable, 
%we calculate the corresponding eigenvalues.
As the interaction matrix \eqref{eq:full_ints} is a circulant matrix,
its eigenvalues are easily calculated as
\begin{equation}
    \label{eq:tiled_one_eigval}
    \begin{cases}
        \operatorname{Re}(\lambda_\ell) & = - \frac{\alpha \rho_0}{D}\left(\mup{} - \mus{} \right) \left[ 1 - \cos( 2 \pi \ell / M) \right],
        \\
        \operatorname{Im}(\lambda_\ell) & = \frac{\alpha \rho_0}{D}\left( \mus{} + \mup{} \right) \sin ( 2 \pi \ell / M),
    \end{cases}
\end{equation}
%for $\ell \in \{ 1, \ldots, M \}$, %for consistency
%where we have defined the rescaled mobilities $\hat{\mu}^{(\text{s,p})} \equiv  \mu^{(\text{s,p})} \alpha \rho_0 / D$
% (see Supplementary Information %\cite{suppmat} 
for graphical representations of the eigenvalue spectra for different species numbers). There are now $M-1$ nonzero eigenvalues, which come as pairs of complex conjugate numbers with the possible exception of $\lambda_{M/2}$ for $M$ even. In stark contrast with the non-cyclic system, the complex character of these eigenvalues opens the door to oscillatory behaviour. The instability condition, obtained by imposing that the real part of at least one eigenvalue is larger than zero, in turn corresponds to
\begin{equation}
    \label{eq:tiled_instab_cond}
     \mup{} - \mus{}< 0,
\end{equation}
i.e.~the catalytic species have to be self-attracting for an instability to occur. This is represented in the phase diagrams of \cref{fig:phase_diag}: all interaction networks above the green line are unstable. If the condition is not satisfied, the system remains homogeneous, with several possible states: the particles can form transient self-propelled molecules 
%(Extended Data Fig. 2(a), Movie 1, see Supplementary Information for the parameters of all Movies and the Extended data),
or form more long-lived, rotating molecules 
%(Extended Data Fig. 2(b), Movie 2)
which exchange particles without growing, as found in particle-based Brownian dynamics simulations of the same system. %(Supplementary Information).

Remarkably, we find key differences between cycles with even or odd number of species. In the case of an even species number $M=2 K$, the eigenvalue with largest real part (which dominates the instability) is real and given by 
%\cite{suppmat}
\begin{equation}
    \label{eq:min_eig_even}
        \lambda_K = - 2 \,  \frac{\alpha \rho_0}{D} \left(\mup - \mus\right) ,
\end{equation}
%which is real, 
implying that the instability is non-oscillatory with the corresponding eigenvector
\begin{equation}
    \label{eq:evec_even}
    \delta\rho^{K}
    %\left(\lambda_K\right)
    =\left( 1, -1, 1, -1, \cdots, -1 \right).
\end{equation}
Thus, at onset of instability, all the species with equal parity tend to aggregate together and to separate from the species of opposite parity (\cref{fig:phase_diag}(a), above the green line). Brownian dynamics simulations show that this prediction carries over to the final phase-separated state; an example is shown in \cref{fig:even_odd_aggreg}(a).
%(Movie 3). 
These simulations show an initial exponential growth of $M$ clusters, each containing all the particles of a given species. The steady state for an even number of self-attracting, cross-repelling species is two large ``clusters of clusters'', one encompassing clusters of the even-labelled species, the other of the odd-labelled species. Both the transient and the steady state are captured by the growth dynamics shown in \cref{fig:parts_cycles}(e), with the average cluster size initially growing exponentially and saturating at half of the total particle population.

A variety of behaviour is observed in the case with chasing interactions among neighbours, based on the relative values of the chasing strength $|\mus+\mup|$ as compared to the self-attraction strength $|\mup-\mus|$. If both values are of the same order of magnitude, the system behaves similarly to the cross-repelling case, except that the resulting clusters can chase each other or rotate in place. % (Extended Data Fig. 1(a) and Movie 4).
For larger inclusions, 
the resulting hybrid clusters indefinitely chase each other,
with transient pairing of same-parity species. %(Movie 6)
For the cases where the value of the self-attraction is much lower than the chasing strength, fully-hybrid clusters containing all species of the same parity form over longer timescales, as opposed to ``clusters of clusters'' as in the cross-repelling case. % (Movie 5).
Finally, for almost negligible self-attraction
%Can we be more specific? Are we saying almost negligible self attraction or what?
transient oscillations are observed before cluster formation. % (Extended Data Fig. 2(b), Movie 6).

\begin{figure}[t]
%   \centering
    \includegraphics[width=1\linewidth]{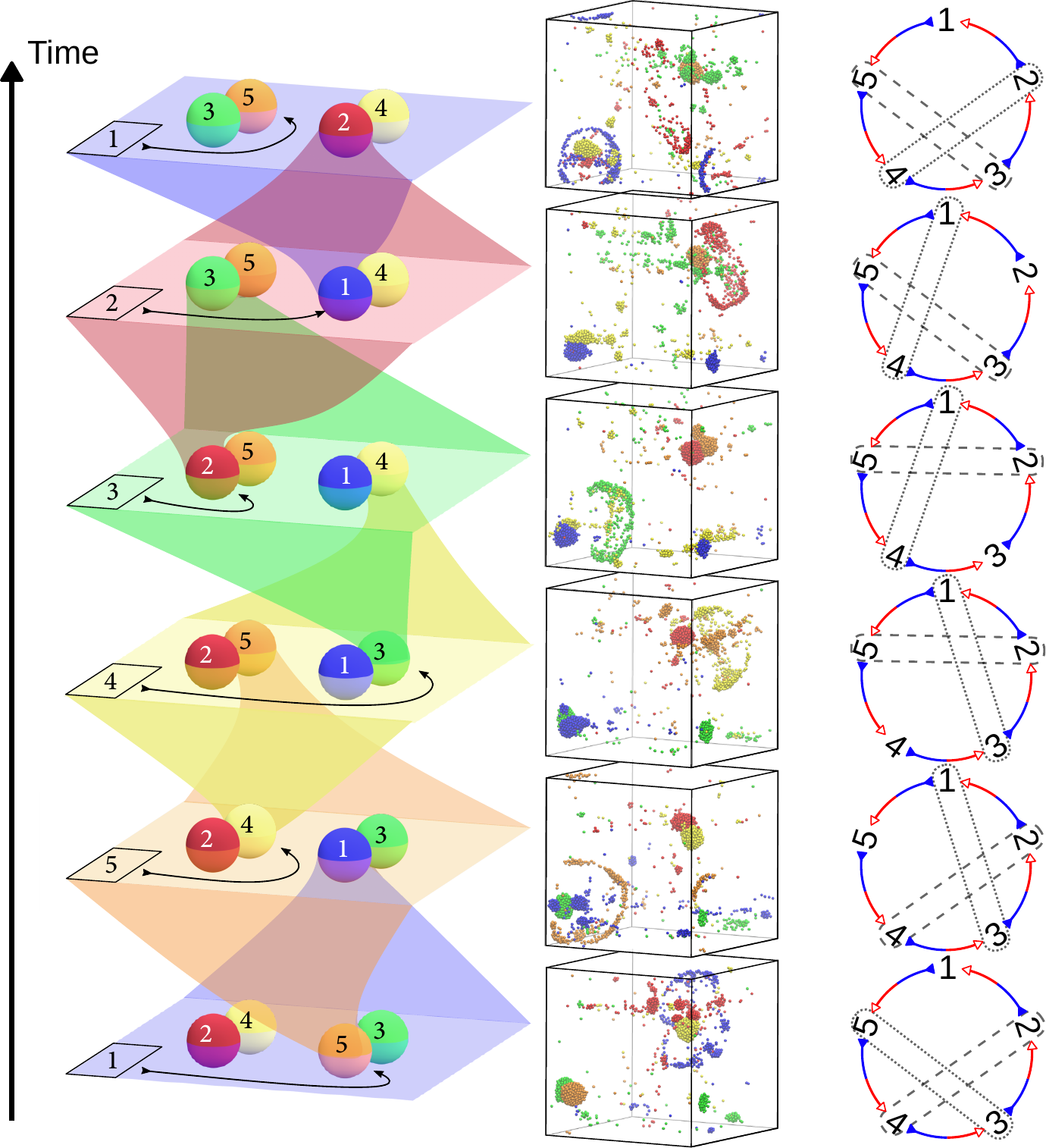}
    \caption{
        {Oscillatory dynamics for an odd number of species with chasing cross-interactions}. A schematic representation of the oscillatory dynamics (left), snapshots of molecular dynamics simulations (middle), %, see also Movie 8), 
        and a diagram of the corresponding species interactions and pairing (right) are shown side by side. Here, dashed and dotted lines represent respectively the pairs on the left and right of the schematic representation. The eigenvalues of the system are as in \cref{fig:even_odd_aggreg}(b), but now $\mathrm{Re}(\lambda)<\mathrm{Im}(\lambda)$ for the most unstable conjugate pair, so that the dynamics of the system are oscillatory.
}
    \label{fig:odd_oscs}
\end{figure}

For cycles with an odd number of species $M=2K+1$, the largest real part corresponds to the complex conjugate pair of eigenvalues %(see Supplementary Information)
\begin{equation}
    \label{eq:min_eig_odd}
    \begin{aligned}
        \lambda_{K+\frac{1}{2} \pm \frac{1}{2}} = &- \frac{\alpha \rho_0}{D} \left(\mup - \mus\right) \left[ 1 + \cos\left(\frac{\pi}{2 K+1}\right) \right] \\
        &\mp i \, \frac{\alpha \rho_0}{D} \left(\mus{} + \mup{}\right) \sin\left(\frac{\pi}{2 K+1}\right),
    \end{aligned}
\end{equation}
suggesting the potential for long-lived oscillations, or even oscillatory steady states, with the real part corresponding to the growth rate of the perturbation and the imaginary part to its oscillation frequency.

The corresponding eigenvectors $\delta\rho^{K+\frac{1}{2}\pm \frac{1}{2}}$ %\equiv \delta \rho_\pm$ 
are also a pair of complex conjugates, with components given by
\begin{equation}
\delta\rho^{K+\frac{1}{2}\pm \frac{1}{2}}_m= (-1)^{m-1}  \left[ \cos{\left(\frac{(m-1) \pi}{2 K+1}\right)} \pm i \sin{\left(\frac{(m-1) \pi}{2 K+1}\right)} \right],
%	\delta\rho_{\pm,j}
\end{equation}
for $m=1,...,2K+1$. The species are out of phase by $2\pi/(2K+1)$ with respect to their second-nearest neighbour during the oscillations.
%From the imaginary part of the eigenvectors, we uncover that the species are out of phase with respect to each other during the oscillations by $2\pi/(2 K+1)$. % of the oscillation period. % and occupy the same location at different times. 
Since the number of species is odd, parity-based cluster aggregation is not possible: if two clusters attempt to come together, a third will systematically come to break them apart. 
%The net result is the formation of pure clusters which separate.
For cross-repelling species, this leads to a segregation into single-species clusters which separate in a way that minimizes their overall repulsion (\cref{fig:even_odd_aggreg}(b)). %, Movie 7).
Similarly to the  even case with $ M = 2K $, this behavior is captured by the growth statistics displayed in \cref{fig:parts_cycles}(e), where mean cluster size exhibits an initial
exponential growth and saturates at a value corresponding to the formation of $M$ individual clusters.

In the case of chasing cross-interactions, oscillations become visible when the growth rate is slower than the oscillation frequency, which corresponds to the condition
\begin{equation}
    \label{eq:tiled_re_vs_im}
    -\mup{} \lesssim
    -\mus{}
   \left[ \frac{1  + \cos\left(\frac{\pi}{2K+1}\right) \mp \sin\left(\frac{\pi}{2K+1}\right)
        }{1  + \cos\left(\frac{\pi}{2K+1}\right) \pm \sin\left(\frac{\pi}{2K+1}\right)  }\right],
\end{equation}
which defines the orange region in \cref{fig:phase_diag}(b). We note that this inequality only sets an order of magnitude for the transition from oscillatory to non-oscillatory dynamics, rather than a sharp boundary. The behaviour of the system again depends on the relative values of the self attraction magnitude $\left| \mup{} - \mus{} \right|$ and the chasing strength $\left| \mup{} + \mus{} \right|$. When self-attraction is weaker than the chasing strength (i.e. close to the instability line), Brownian dynamics simulations indeed show a persistent %extremely long-lived 
oscillatory dynamical behaviour 
%(likely permanent, as they persist unchanged for the duration of our simulations), 
with the following choreography for the case in which each species chases after the previous one: a single-species cluster of a species $m$ forms transiently, and is then ``invaded'' by species $m+1$, leading to an explosion that disperses species $m$ back into the solution. Species $m$ then invades a cluster of species $m-1$, and so on, in a sequential order until $M$ explosion events have occurred and the cycle starts again. In the case with $M=5$ (\cref{fig:odd_oscs}), % and Movie 8; see Supplementary Information for a quantification of the oscillation dynamics), 
we observe that the system comes back to a state similar to the initial one, except for a swap in the locations of the clusters. This change occurs because the clusters of the second-nearest-neighbour species in the cycle tend to form pairs. One component of one of these pairs is replaced in every explosion event by the species preceding it in the cycle, such that, after five explosions, the pairs have been switched in space.
The reverse dynamics (species $m$ invading species $m+1$) are observed if the signs of $\mus$ and $\mup$ are reversed, so that each species chases the next one in the cycle.

%The mechanism for these oscillations is linked to strong chasing compared to self-attraction. In the first simulation snapshot in \cref{fig:odd_oscs}, species 5 has just formed a cluster, towards which active particles of species 1 start converging due to their strong attraction to species 5. The strong repulsion that species 1 exerts on species 5 then violently breaks up the cluster of the latter. This leads to the replacement of the species 5 cluster with a species 1 cluster, while the species 5 particles are dispersed in the simulation box. This process is then repeated, with the now free particles of species 5 attacking the cluster of species 4 and replacing it, dispersing species 4 particles. After five such explosion events, the system comes back to a state similar to the initial one, except for a swap in the locations of the clusters. This change occurs because the clusters of the second-nearest-neighbour species in the cycle tend to form pairs. One component of one of these pairs is replaced every explosion event by the species preceding it in the cycle, such that, after five explosions, the pairs have been switched in space.

For even weaker self-attraction or stronger chasing, the clusters do not have time to form. In this case, oscillations are observed in a dilute mixture of catalytic particles, where clusters are replaced by transient zones of higher concentration. %(Movie 9).
This can create a self-stirring solution, favouring the mixing and assembly of solution components in time scales considerably shorter than those allowed by passive diffusion. Lastly, if the perturbation growth rate is instead larger than its oscillation frequency (red region in \cref{fig:phase_diag}(b)), then the dynamics leads to formation of stable clusters. We have observed in simulations the formation of chasing hybrid clusters similar to the case with even number of species. % (Extended Data Fig. 3, Movie 10).

Our work shows that catalytically-active and chemotactic particles participating in a primitive metabolic cycle exhibit a variety of structural complex collective behaviour. Due to the nature of the gradient-mediated interactions involved, such particles are able to interact over large distances, and undergo spontaneous and exponentially rapid cluster formation that serves to support their metabolic function. Depending on the parity of the number of different species involved in the cycle and on their chemotactic parameters, these clusters might consist of a single or several species, thereby accommodating a range of design strategies for metabolic structure formation. If the number of species in the cycle is odd, chasing interactions may emerge at the macroscopic level, similar to those that have been observed in recent experiments \cite{qiao2017Predatory,meredith2020Predator}, although in this case leading to long-lived, system-wide oscillations. The observed variety of emergent structural behaviour with highly precise control over the composition of the constituents of the metabolically active clusters hints at a significant possible role for catalytically active molecules at the origin of life: the molecules that are metabolically connected to each other will preferentially and efficiently form active clusters together, hence serving as potential candidates for the nucleation of early forms of life. 
%metabolon \cite{pareek2021metabolic}

\acknowledgements
We acknowledge support from the Max Planck School Matter to Life and the MaxSynBio Consortium which are jointly funded by the Federal Ministry of Education and Research (BMBF) of Germany and the Max Planck Society.

% \section*{Author Contributions Statement} 

% V.O-.R., J.A-.C., and R.G. designed the research, conducted the research, analyzed the data, and wrote the paper. %other contributions to be highlighted...

% \section*{Competing Interests Statement} 

% The authors declare no competing financial interests.

\FloatBarrier

\bibliography{mybib-v3}   % name your BibTeX data base
\FloatBarrier

\clearpage

% \section*{Data Availability} 

% The data supporting the main findings of this study are available in the paper %and its Supplementary Information.
% Any additional data can be made available upon request.

% \section*{Code Availability} 

% The algorithms for the codes supporting the main findings of this study are available in the paper %and its Supplementary Information.
% Any additional information concerning the code can be made available upon request.

\end{document}